\documentclass[aps,pra,twocolumn,showpacs]{revtex4}
\usepackage{color, graphicx}
\usepackage{amsmath, amssymb}
\usepackage{ulem,hyperref}

\usepackage{tikz} 
\usetikzlibrary{fit,matrix} 
\usepackage{braket} 

\begin{document}

\title{Entanglement and phase properties of noisy N00N states}

\author{M. Bohmann}\email{martin.bohmann@uni-rostock.de}\affiliation{Arbeitsgruppe Theoretische Quantenoptik, Institut f\"ur Physik, Universit\"at Rostock, D-18051 Rostock, Germany}
\author{J. Sperling}\affiliation{Arbeitsgruppe Theoretische Quantenoptik, Institut f\"ur Physik, Universit\"at Rostock, D-18051 Rostock, Germany}
\author{W. Vogel}\affiliation{Arbeitsgruppe Theoretische Quantenoptik, Institut f\"ur Physik, Universit\"at Rostock, D-18051 Rostock, Germany}

\pacs{03.67.Mn, 42.50.Ex, 42.50.St, 03.67.Hk}

\begin{abstract}
	Quantum metrology and quantum information necessitate a profound study of suitable states.
	Attenuations induced by free-space communication links or fluctuations in the generation of such states limit the quantum enhancement in possible applications.
	For this reason we investigate quantum features of mixtures of so-called N00N states propagating in atmospheric channels.
	First, we show that noisy N00N states can still yield a phase resolution beyond classical limitations.
	Second, we identify entanglement of noisy N00N states after propagation in fluctuating loss channels.
	To do so, we apply the partial transposition criterion.
	Our theoretical analysis formulates explicit bounds which are indispensable for experimental verification of quantum entanglement and applications in quantum metrology.
\end{abstract}

\date{\today}

\maketitle

\section{Introduction}
	Quantum mechanics has opened new fields of applications such as quantum metrology~\cite{Giovannetti2004,Giovannetti2011}, quantum information science~\cite{Nielsen}, and quantum communication~\cite{Gisin2007}.
	All these technologies have in common that they employ nonclassical properties such as entanglement in order to surpass classical restrictions.
	In quantum metrology such a classical limitation is the so-called standard quantum limit or shot noise limit~\cite{Giovannetti2004,Giovannetti2011}.
	It relates the  error of the estimation $\Delta\varphi$ of a parameter $\varphi$, e.g., the phase, with the energy or the photon number.
	Employment of classical strategies leads only to the bound $\Delta\varphi\gtrsim1/\sqrt{n}$, where $n$ denotes the mean photon number.
	However, using nonclassical states in detection strategies this behavior can be beaten; cf., e.g.,~\cite{SubSQL1,SubSQL2,SubSQL3}.
	High-precision quantum interferometric measurements allow a resolution up to the Heisenberg limit $\Delta\varphi\sim1/n$~\cite{Giovannetti2004}.
	Superior phase estimations are also used to answer open questions in physics, e.g., gravitational wave detection~\cite{Caves,Gravi}.

	One prominent representative of quantum states which reaches the Heisenberg limit is the two-mode, quantum entangled N00N state~\cite{N00N}.
	The name originates from its expansion in Fock bases:
	\begin{equation}\label{gl:N00N}
		|\psi_N\rangle=\frac{1}{\sqrt{2}}\left(|N\rangle\otimes |0\rangle+|0\rangle\otimes |N\rangle\right).
	\end{equation}
	N00N states are Bell-like entangled states.
	Besides the application in quantum metrology~\cite{Giovannetti2011} they are also of interest in quantum lithography~\cite{N00N}.
	By employing N00N states, a phase supersensitivity --  i.e., a phase estimation better than any classical estimation -- in a Mach-Zehnder interferometer can be achieved.
	Several strategies to generate such states have been proposed and experimentally implemented; see, e.g.,~\cite{N00NG1,N00NG2} and~\cite{N00NG3,N00NG4}, respectively.

	In addition to quantum metrology, secure communication based on quantum key distribution (QKD) is another vital topic of current research~\cite{Gisin2007}.
	Although QKD is already used commercially, it is still restricted to relatively small transmission range.
	In systems using optical fibers the loss has a typical magnitude of $\sim0.2$ \text{dB}/\text{km}~\cite{Mitschke}, which renders it possible to preserve entanglement up to a $\sim100\,\text{km}$ propagation distance~\cite{Fibre1,Fibre2,Fibre3}.
	An alternative approach is the quantum communication through free-space links.
	QKD in such links was demonstrated for an atmospheric channel of $144$ \text{km}~\cite{Ursin,Fedrizzi} and recent experiments show the feasibility of global communication via orbiting satellites; see, e.g.,~\cite{Satellite0,Satellite1,Satellite2}.
	It is noteworthy that the implemented protocols use postselection strategies.
	In order to benefit from free-space QKD it is crucial to study the turbulent losses occurring within such a link.
	The first consistent quantum description of turbulent loss channels was introduced in terms of fluctuating transmission coefficients~\cite{Semenov2009}.

	QKD employs entangled states as a resource.
	One of the most frequently used methods for determining this kind of quantum correlation in the bipartite case is formulated in terms of negativity of the partial transposition (PT) of the density operator.
	In this way entanglement is tested by transposing one of the involved modes while leaving the other one unchanged.
	If this partially transposed state cannot fulfill the positivity required for quantum states the considered quantum system is entangled~\cite{Peres,Horodecki1996}.

	In this contribution we study mixtures of N00N states propagating in atmospheric channels.
	We show that the superior phase-estimation properties can surpass perturbations within interferometric setups without postselection.
	Entanglement is tested by the partial transpose criterion.
	In particular, the entanglement properties are studied in the case of turbulent losses occurring in urban free-space links.
	Surprisingly simple conditions are deduced to infer sub-standard quantum limit behavior of phase estimation.
	These conditions may be used to estimate rough bounds to experimentally allowed ranges of noise to infer such quantum features.

	In Sec.~\ref{ch:mixedN00Nstates} we motivate the generalization from pure N00N states to noisy ones.
	We study the phase properties of this class of states within an interferometric setup in Sec.~\ref{ch:Phase}.
	The entanglement and phase properties are further studied under turbulent loss conditions in Sec.~\ref{ch:App}.
	A summary and conclusions are given in Sec.~\ref{ch:Summary}.

\section{Noisy N00N states}
\label{ch:mixedN00Nstates}
	In this section we motivate the class of noisy N00N states as a generalization of ordinary N00N states.
	Such a generalization is natural, considering a possible scheme of generating N00N states.
	Moreover we consider the effects of atmospheric channels as a source of turbulent losses which may alter the correlation properties of the states under study.
	Another source of possible imperfections is dephasing, which will be investigated subsequently.
	Eventually, we will formulate a general family of quantum states which are based on mixed N00N states including all previously considered perturbations.

\subsection{Generation of mixed N00N states}
	Here we will focus on one previously implemented generation of N00N states by using coherent and squeezed vacuum states at the input ports of a symmetric beam splitter.
	Such a generation strategy has been studied in theory~\cite{Hofmann,Pezze} and has been successfully applied in experiments~\cite{Afek,Israel}.
	For example, any path entangled N00N state for any $N$ can in principle be generated in this form with a fidelity of at least $92\%$~\cite{Hofmann}.
	In the experiments conducted so far, e.g.,~\cite{Afek}, this generation strategy was used in Mach-Zehnder interferometers with the aim of surpassing the standard quantum limit.
	Via coincidence measurements the $N$-fold super-resolution was demonstrated for up to $N=5$.
	Analyzing the two mode photon number statistics of the state in the interferometer, one observes that the generated state is a mixture of N00N states rather than a pure single N00N state.

	This fact motivates the study of mixed N00N states,
	\begin{equation}\label{gl:mixedN00N}
	\hat{\rho}_{\text{mixed}}=\sum\limits_{N=0}^\infty p_N|\psi_N \rangle\langle \psi_N|
	\end{equation}
	with $p_N$ being a probability distribution of N00N states $|\psi_N \rangle$ in Eq.~\eqref{gl:N00N}.
	For $N=0$ we use the convention $|\psi_0 \rangle=|0,0\rangle$, which is the two-mode vacuum state.
	These mixtures properly describe the experimentally realized states.

\subsection{Turbulent loss channels}\label{ch:ModelAt}\label{ch:TurMixN00N}
	A crucial issue using N00N states is the sensitivity with regards to loss.
	One imagines a loss of a photon in one of the modes--say the second mode.
	In this case the entangled N00N state~(\ref{gl:N00N}) reduces to a non-entangled state, $|\psi_N\rangle\mapsto|0,N-1\rangle$.
	The outcome loses all the rewarding properties of the N00N state and especially its supersensitivity in phase measurement.
	Thus it is of great interest i studying the quantum properties of mixed N00N states undergoing losses.
	Note that the influence of noise effects on metrology scenarios has been studied generally in Refs.~\cite{NoisyMetrology1,NoisyMetrology2,NoisyMetrology3}.

	For a single N00N state propagating in a constant loss channel $\Lambda_{\kappa,\theta}$, we get
	\begin{align}\label{gl:lossy}
		\nonumber &\Lambda_{\kappa,\theta}(|\psi_N\rangle\langle\psi_N|)\\
		\nonumber =&\frac{\sqrt{\kappa\theta}^N}{2}\Big[
		|N,0\rangle\langle 0,N|+|0,N\rangle \langle N,0|\Big]\\
		\nonumber &+\frac{1}{2}\sum\limits_{k=0}^N\binom{N}{k}\Big[ \kappa^k(1-\kappa)^{N-k}|k,0\rangle\langle k,0| \\
		&\phantom{+\frac{1}{2}\sum\limits_{k=0}^N\binom{N}{k}+} +\theta^k(1-\theta)^{N-k}|0,k\rangle\langle 0,k|\Big],
	\end{align}
	where the transmission coefficients $\sqrt{\kappa}$ and $\sqrt{\theta}$ describe the losses, $1-\kappa$ and $1-\theta$, in the first and second subsystems, respectively.
	Starting from~(\ref{gl:lossy}) the turbulent N00N states will be derived, representing attenuated states after propagation through a fluctuating loss medium.

	The turbulent atmosphere differs from standard constant loss channels due to the fact that the transmission coefficient is not a constant but a random variable.
	In this way the description of turbulence is included in the probability distribution of the transmission coefficient (PDTC)~\cite{Semenov2009}, denoted as $\mathcal{P}$.
	The density operator is obtained by averaging the lossy state with an appropriate PDTC.
	In a bipartite system, this process reads as
	\begin{align}\label{eq:PDTCchannel}
		\hat\rho_{\rm out}=\int\limits_0^{1}\int\limits_0^{1}d\sqrt{\kappa}\,d\sqrt{\theta}\,
		\mathcal{P}(\sqrt{\kappa},\sqrt{\theta})\,\Lambda_{\kappa,\theta}(\hat\rho_{\rm in}).
	\end{align}
	In the following we want to focus on the effect of beam wandering because of its well characterized PDTC~\cite{beamwandering} and due to its importance for urban communication links.
	It has been demonstrated for this scenario, that theory and experiment show excellent agreement with each other~\cite{Usenko}.
	For a single spatial-mode, the PDTC can be expressed as a log-negative Weibull distribution,
	\begin{equation}\label{gl:tur}
		\mathcal{P}(T)=\frac{2S^2}{\sigma^2\zeta T}\left(2\ln\frac{T_0}{T}\right)^{\!\!\frac{2}{\zeta}-1}\exp\!\left[-\frac{1}{2\sigma^2}S^2\left(2\ln\frac{T_0}{T}\right)^{\!\!\frac{2}{\zeta}}\right],
	\end{equation}
	for $T\in[0,T_0]$ and $\mathcal{P}(T)=0$ else.
	The meaning of the occurring constants are listed in Appendix~\ref{app:Weinbull}.
	As an example of an atmospheric channel we will consider a quantum channel in an urban environment as in experiments performed in Erlangen~\cite{Elser,Heim} and theoretically studied in~\cite{Semenov2012}.
	Note that there have also been other experiments performed in this urban quantum free-space channel regime, e.g.,~\cite{Usenko,Aspelmeyer,Peuntinger}.

	In order to describe the state after propagation through the atmosphere, see Eq.~\eqref{eq:PDTCchannel}, we will now apply the PDTC model to the lossy N00N state~(\ref{gl:lossy}).
	This can be done in two configurations.
	First, the two modes could propagate in different directions yielding a joint PDTC in terms of independent single mode ones,
	\begin{align}\label{eq:counter-prop}
		\mathcal P(\sqrt\kappa,\sqrt\theta)=\mathcal P(\sqrt\kappa)\mathcal P(\sqrt\theta).
	\end{align}
	Second, a co-propagation of the field components could be considered.
	According to the experiments in Erlangen we might consider the polarization as the two degrees of freedom.
	In this case the two modes undergo the same turbulence, i.e. the joint PDTC reads as
	\begin{align}\label{eq:co-prop}
		\mathcal P(\sqrt\kappa,\sqrt\theta)=\mathcal P(\sqrt\kappa)\delta\left(\sqrt\kappa-\sqrt\theta\right),
	\end{align}
	with $\delta$ being the Dirac delta distribution.

\subsection{Dephasing and the general state representation}
	Among other perturbations, dephasing is a harmful influence on the phase sensitivity of N00N states.
	A dephasing single-mode channel can be described by
	\begin{align}
		\hat\rho_{\rm out}=\Lambda_{\rm deph}(\hat\rho_{\rm in})=\int_{0}^{2\pi} d\varphi\, p(\varphi) e^{i\varphi\hat n}\hat\rho_{\rm in}e^{-i\varphi\hat n},
	\end{align}
	for a classical phase distribution $p(\varphi)$ and with $\hat n$ being the photon number operator.
	In the case of the N00N state, we get for the dephasing in one mode
	\begin{align}
		\nonumber &\mathbb I\otimes\Lambda_{\rm deph}(|\psi_N\rangle\langle\psi_N|)\\
		\nonumber =&\frac{1}{2}\left[|0,N\rangle\langle0,N|+|N,0\rangle\langle N,0|\right.\\
		&\phantom{\frac{1}{2}}\left.+\lambda|N,0\rangle\langle 0,N|+\lambda^\ast|0,N \rangle\langle N,0|\right],\label{gl:GaussN00N}
	\end{align}
	with $\lambda=\int_{0}^{2\pi} d\varphi\, p(\varphi) e^{i\varphi N}$.
	Note that, in general, we have reduction of coherence, $|\lambda|<1$.
	Moreover, full decoherence, $p(\varphi)=1/(2\pi)$, yields $\lambda=0$.
	That is a complete loss of phase sensitivity and entanglement.

	Finally we combine all imperfections considered so far which yields a very general class of noisy NOON states.
	The resulting density matrix in a Fock expansion reads as
	\begin{align}
		\nonumber \hat{\rho}=&\rho_{00,00} |0,0\rangle \langle 0,0|\\
		\nonumber &+\sum\limits_{i=1}^{\infty} \Big[\rho_{i0,i0} |i,0\rangle \langle i,0|+\rho_{0i,0i} |0,i\rangle \langle 0,i|\Big]\\
		&+\sum\limits_{i=1}^{\infty}\Big[\rho_{i0,0i} |i,0\rangle \langle 0,i|+\rho_{0i,i0} |0,i\rangle \langle i,0|\Big].\label{eq:NoisyN00N}
	\end{align}
	In this form the first line represents the vacuum contribution to the state being separable and carrying no phase information.
	The second line includes the diagonal entries of $\hat\rho$.
	The last line contains the interferences of the state.
	Because $\hat\rho=\hat\rho^\dagger$, we have $\rho_{i0,0i}={\rho_{0i,i0}}^\ast$.
	From the metrology as well as the entanglement point of view, these terms are those which classify the applicability for quantum enhanced tasks.

	This  general form~\eqref{eq:NoisyN00N} includes mixtures of N00N states, the propagation effects in turbulent media, and dephasing.
	In the following, states of the structure~\eqref{eq:NoisyN00N} will be referred to as noisy N00N states.
	In the remainder of this contribution we characterize such states in terms of phase resolution and entanglement.

\section{Phase properties of noisy N00N states}\label{ch:Phase}
	As we discussed above N00N states are of great interest due to their sub standard quantum limit behavior in interferometric setups.
	We will now show that noisy N00N states can still exhibit this property and give analytical conditions for its appearance.
	Therefore we consider the typically studied measurement that is given by the operator
	\begin{equation}\label{gl:mixedN00Nop}
		\hat{A}_M=|0,M\rangle \langle M,0| +|M,0\rangle \langle 0,M| \, ,
	\end{equation}
	which can be realized via interference measurements at the output ports of a Mach-Zehnder interferometer.
	The error in phase estimation can be calculated via error propagation as
	\begin{equation}\label{gl:Errorpropagation}
		\Delta\varphi=\frac{\sqrt{\langle (\Delta \hat{A}_M)^2\rangle}}{\frac{d|\langle \hat{A}_M \rangle|}{d\varphi}}.
	\end{equation}
	Computing the expectation values we get a phase-dependent error
	\begin{align}\label{gl:phasemixedN00N}
		\Delta\varphi=&\frac{1}{M}\left[\frac{\rho_{M0,M0}+\rho_{0M,0M}-\left[2{\rm Re}(e^{i\varphi M}\rho_{0M,M0})\right]^2}{\left[2{\rm Im}(e^{i\varphi M}\rho_{0M,M0})\right]^2}\right]^{\frac{1}{2}}\nonumber\\
		=&\frac{f_M(\varphi)}{M}
	\end{align}
	This expression still includes the $1/M$ behavior as obtained for a perfect N00N state.
	However, we have an additional phase depends scaling factor $f_M(\varphi)$ which also dependents on $M$.

	Due to the fact that the function $f_M(\varphi)$ in \eqref{gl:phasemixedN00N} depends on both the phase $\varphi$ itself and $M$ given by the addressed interference term \eqref{gl:mixedN00Nop} one preserves only certain intervals in which phase supersensitivity is observable.
	Note that especially outside these intervals the phase error is worse than the standard quantum limit, cf. Fig.~\ref{fig:ErrormixedN00N}.
	The phase interval in which the standard quantum limit is surpassed, $\Delta\varphi<1/\sqrt M$, can be shown to be determined by 
	\begin{align}\label{gl:mixedN00Ninterval}
		\frac{1}{M}\arcsin\left(\omega\right)&<\varphi+\frac{\arg(\rho_{0M,M0})}{M}<\frac{\pi}{M} -\frac{1}{M}\arcsin\left(\omega\right),\\
		\text{ 	with } \omega&=\sqrt{\frac{\rho_{M0,M0}+\rho_{0M,0M}-4|\rho_{0M,M0}|^2}{4|\rho_{0M,M0}|^2(M-1)}}.\nonumber
	\end{align}
	Note that this interval has a $\pi/M$ periodicity.
	In order to have solutions for the inverse sine function, an additional condition has to be fulfilled:
	\begin{align}\label{gl:pM}
		|\rho_{0M,M0}|^2>\frac{1}{M}\frac{\rho_{M0,M0}+\rho_{0M,0M}}{4}.
	\end{align}

	In order to achieve phase supersensitivity for such an estimation scheme, one needs to have some prior knowledge of the actual phase value.
	Otherwise the average estimation procedure performs worse than the standard quantum limit.
	This is due to the phase regions violating Eq.~\eqref{gl:mixedN00Ninterval}, where $\Delta\varphi$ exceeds the standard quantum limit; also see Fig.~\ref{fig:ErrormixedN00N}.
	Nevertheless one can always get some prior information about $\varphi$.
	Here one can imagine two different scenarios.
	In the first case one wants to measure small deviations from a given, known phase value.
	Then the prior phase information is precisely this given phase value and thus it is directly at hand.
	In the second case, prior phase information can be obtained by inspecting the expectation value $\langle\hat A_M\rangle$ of the observable~\eqref{gl:mixedN00Nop}.
	This expectation value oscillates with the phase and its roots coincide with the points of minimal phase estimation error, in our case an odd multiple of $\varphi=\pi/M$.
	These phase points correspond to destructive interference and thus are experimentally accessible.
	Note that such an approach is well known; see, e.g.,~\cite{Dowling,Kim}.
	In this case, a small fraction of the available copies of the used state will be employed to roughly estimate the prior phase value.
	Consequently, this procedure consumes a small part of the resources of the estimation process~\cite{Prior,Prior1}.
	For a more general treatment in the global approach, without {\it a priori} phase knowledge, we refer to~\cite{NoPrior}.

\subsection{Mixtures}
\begin{figure}[h]
		\includegraphics[clip,width=1\linewidth]{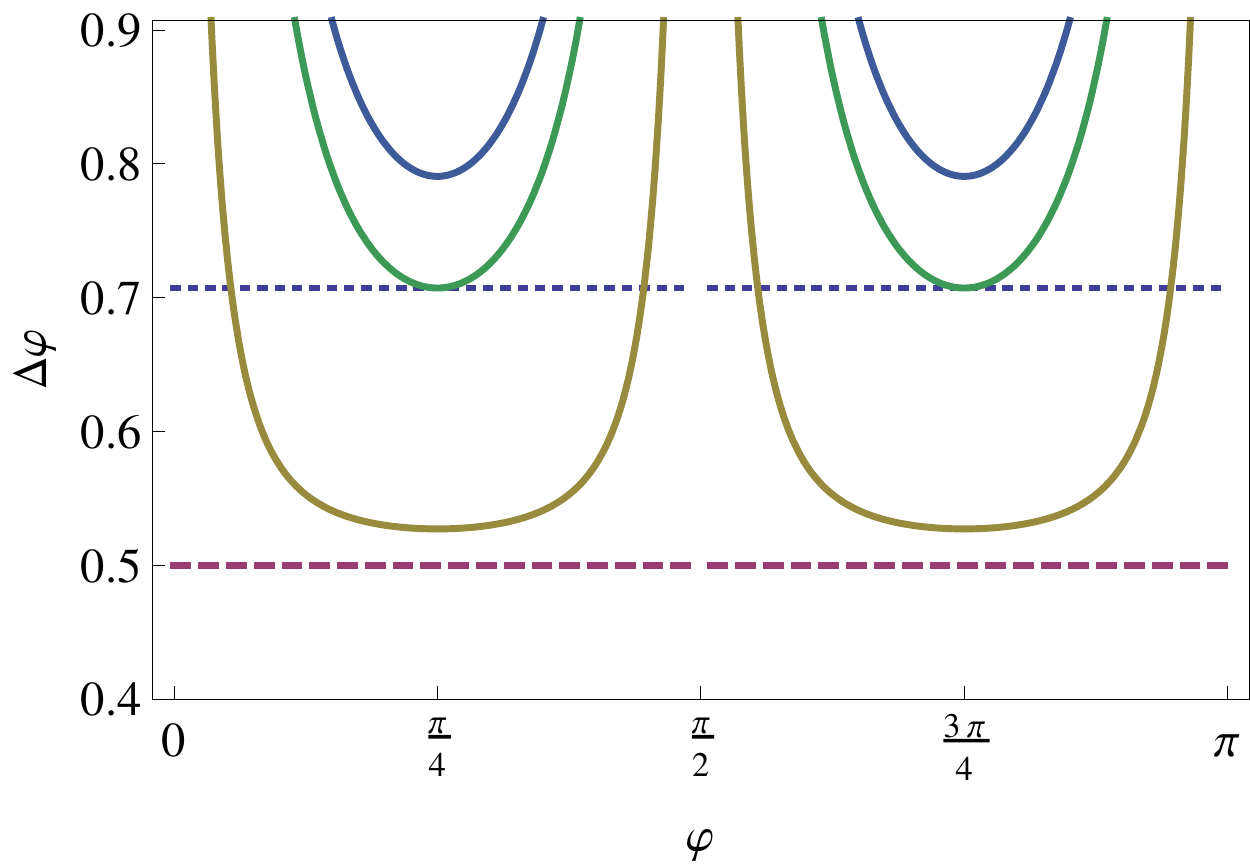}
		\caption{(Color online) 
			Here the $\pi/M$-periodic error in phase estimation $\Delta\varphi$ (solid curves) for different mixing coefficients $p_M=0.4$, $0.5$, and $0.9$ (from top to bottom), is shown for $M=2$; see also Eq.~\eqref{gl:mixedN00N}.
			The corresponding limits $1/\sqrt{M}$ (dotted) and $1/M$ (dashed) are given.
		}
		\label{fig:ErrormixedN00N}
	\end{figure}

	For mixtures of N00N states in the form~\eqref{gl:mixedN00N} the phase dependency is illustrated in Fig.~\ref{fig:ErrormixedN00N}.
	Here the error in phase estimation $\Delta\varphi$ for $M=2$ and for different probabilities $p_M$ to realize the $M$th N00N state in an ensemble is displayed.
	Depending on the value of $p_M$, the standard quantum limit can be surpassed.
	In this case Eq.~\eqref{gl:pM} reduces to the rather simple expression $p_M\ge 1/M$.
	This is a remarkable result as it directly relates the quantum fidelity $F$ of mixed N00N states~\eqref{gl:mixedN00N} with a N00N state with $M$ photons.
	Only if $F= p_M\ge 1/M$ can one observe a scaling of $\Delta\varphi$ below the standard quantum limit, compare Fig.~\ref{fig:ErrormixedN00N}.
	Note that in the special case $p_M\to 1$ one reaches the well known $1/M$ behavior of the pure N00N state.

\subsection{Constant loss}
	Let us emphasize at this point that the example discussed above considered mixtures of N00N states~\eqref{gl:mixedN00N} as a special case of noisy N00N states~\eqref{eq:NoisyN00N}.
	Thus phase supersensitivity might not be achievable with every noisy N00N state.
	Especially for lossy N00N states~\eqref{gl:lossy} one meets additional difficulties.
	In this case Eq.~\eqref{gl:phasemixedN00N} reads as
	\begin{align}\label{gl:losserror}
	   \Delta\varphi=\frac{1}{M}\left[\frac{1-\sqrt{\kappa\theta}^M\cos^2 (\varphi M)}{\sqrt{\kappa\theta}^M\sin^2 (\varphi M)}\right]^{\frac{1}{2}}
	\end{align}
	Here the part $f_M(\varphi)$ diverges in the limit of $M\to\infty$ [see Eqs.~\eqref{gl:losserror} and~\eqref{gl:phasemixedN00N}], as the denominator of $f_M(\varphi)$ goes to zero in that limit since $\kappa,\theta\in [0,1[$.
	How fast the quantum enhancement vanishes with respect to $M$ depends on the strength of the loss.  
	We will discuss this fact later in some detail at the end of Sec.~\ref{sec:InfluTur} in the case of atmospheric losses.

\begin{figure}[h]
		\includegraphics[clip,width=1\linewidth]{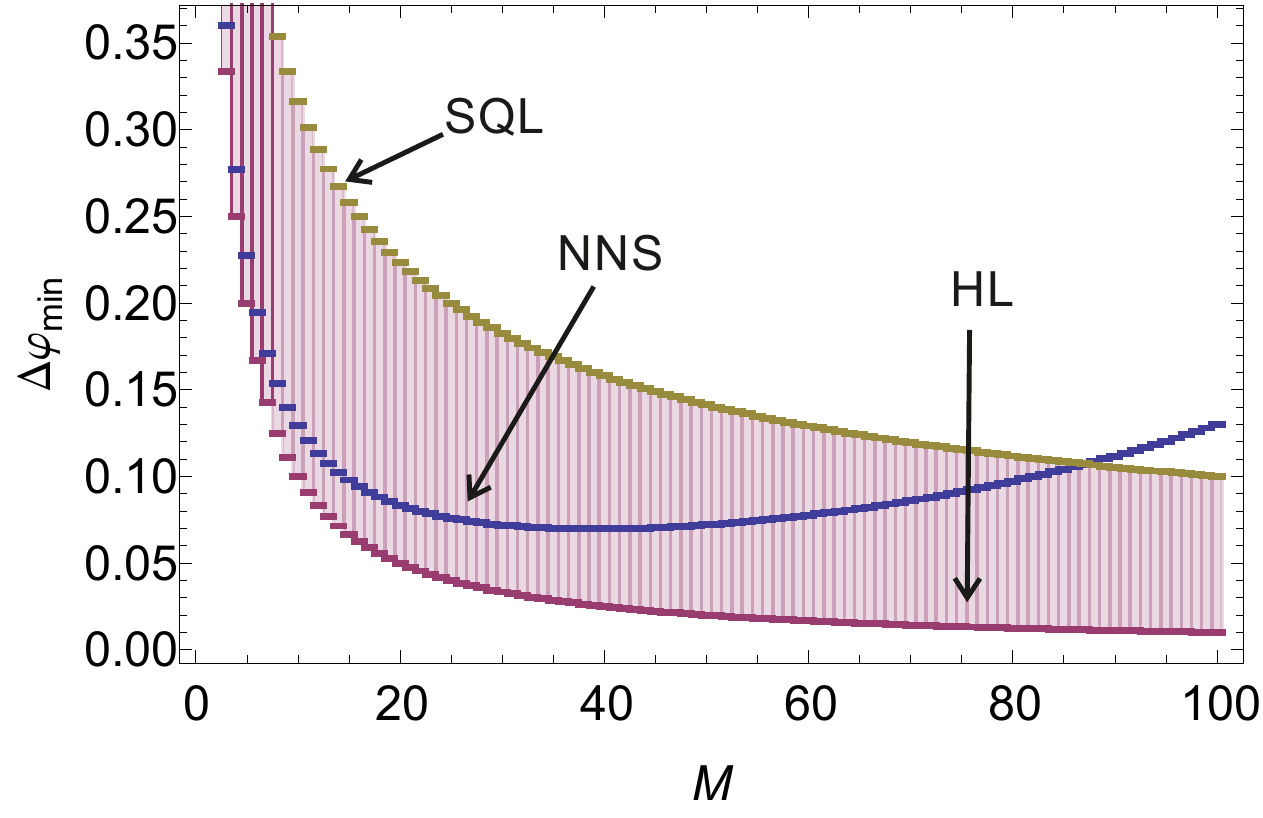}
		\caption{
			(Color online) 
			The minimal error $\Delta\varphi_{\text{min}}$ in phase estimation 
			is displayed over $M$ (blue chart)
			for a noisy N00N state (NNS) with constant loss of $5\%$ in both modes $(\kappa=\theta=0.95)$ .
			Additionally, the $1/M$ Heisenberg limit (HL) scaling  and the $1/\sqrt{M}$ standard quantum limit (SQL) are shown.
			For small values of $M$, $\Delta\varphi_{\text{min}}$ follows the $1/M$ scaling.
			With increasing $M$, the error exceeds the HL but still beats the SQL.
			Eventually, the NNS curve exceeds the SQL at $M=88$.
		}
		\label{fig:ErrorlossyN00N}
	\end{figure}

	For such a constant loss case with equal loss of $5\%$ in both modes $(\kappa=\theta=0.95)$ the $M$-dependency of the minimal error in phase estimation, $\Delta\varphi_{\text{min}}$, is shown in Fig.~\ref{fig:ErrorlossyN00N}.
	One observes that for small values of $M$ the minimal error follows the $\sim 1/M$ Heisenberg scaling, starting to deviate from that scaling with increasing $M$.
	When approaching an optimal value of $M$, given by $M\approx -2/\log\sqrt{\kappa\theta}$ (here: $M=39$), the error decreases with increasing $M$.
	Beyond this point the error increases with larger $M$, exceeds the standard quantum limit at $M=88$ and diverges in the limit of $M\to\infty$.
	However, $\Delta\varphi_{\mathrm{min}}$ is shown in the region below $M=39$ and especially for $M<10$, which is in reach of today's technology, a remarkable scaling close to the Heisenberg limit.
	Hence, a significant improvement with respect to the standard quantum limit can be observed.

	This behavior is consistent with the results given in Ref.~\cite{Referee}, where the estimation precision in the presence of decoherence has been studied in a universal manner.
	By optimization over all possible states it has been shown that the improvement in the phase error in dependence on $M$ asymptotically flattens to a constant factor in the presence of loss.
	Thus the quantum states considered in the present work are not optimal.
	Nevertheless, an advantageous scaling is obtained for small $M$ values for the considered class of noisy N00N states.

\subsection{Dephasing}
	\begin{figure}[h]
		\includegraphics[clip,width=1\linewidth]{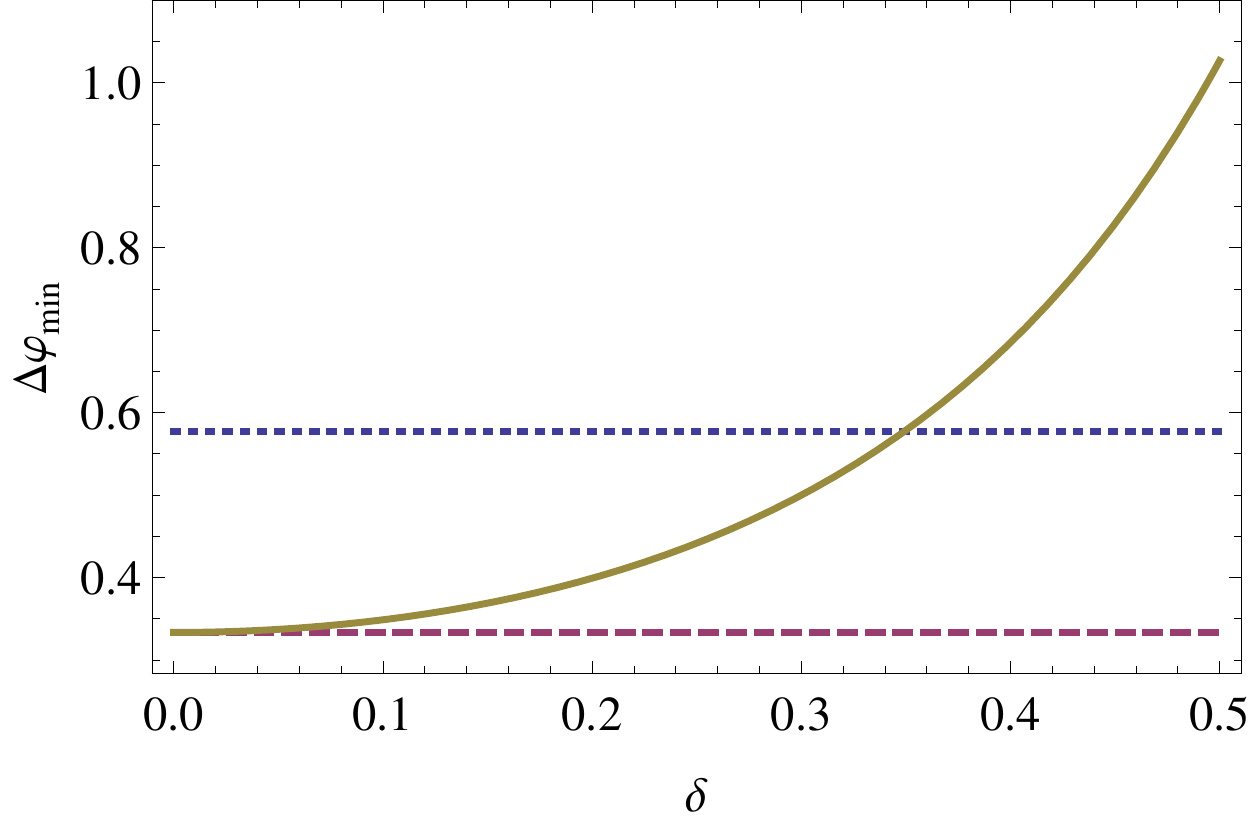}
		\caption{(Color online)
		The error in phase estimation $\Delta\varphi_{\min}$ (solid) for a N00N state with $N=3$ undergoing Gaussian phase noise is plotted in dependence on the noise width parameter $\delta$. 
		Additionally the corresponding standard quantum limit and Heisenberg limit are displayed (dotted and dashed, respectively).
		Even if we have phase noise of $\delta=0.1 \pi$, we observe sub-standard-quantum-limit behavior.
		}
		\label{fig:Dp2}
	\end{figure}
	Second we will study the influence of dephasing on the error in phase resolution.
	Therefore we  will consider a pure N00N state undergoing dephasing [see Eq.~\eqref{gl:GaussN00N}], modeled by a wrapped Gaussian phase distribution in the form
	\begin{align}\label{gl:pGauss}
		p(\varphi)=\sum\limits_{k\in \mathbb Z}\frac{1}{\sqrt{2\pi\delta^2}}\exp\left[-\frac{(\varphi-\varphi_0+2k\pi)^2}{2\delta^2}\right],
	\end{align}
	where $\delta$ controls the strength of the noise.
	Such a phase noise distribution has already been studied in the context of entanglement in Ref.~\cite{Sperling2012}.
	When this is applied to a N00N state, we get
	\begin{align}
	   \lambda=&\int\limits_{0}^{2\pi}d \varphi\, \sum\limits_{k\in \mathbb Z}\frac{1}{\sqrt{2\pi \delta^2}}\exp\left[-\frac{(\varphi-\varphi_0+2k\pi)^2}{2\delta^2}\right] e^{i\varphi N}\\\nonumber
	  &=e^{i\varphi_0N}e^{-\frac{\delta^2N^2}{2}}, 
	\end{align}
	cf. Eq.~\eqref{gl:GaussN00N}.
	The dependency of the error in phase estimation on the noise parameter $\delta$ is shown in Fig.~\ref{fig:Dp2}.
	One directly observes that $\Delta\varphi_{\min}$ increases from the Heisenberg limit, corresponding to $\delta\to 0$, with increasing $\delta$ and even becomes worse than the standard quantum limit for $|\lambda|^2>1/M$, corresponding to
	\begin{align}\label{Eq:PhaseNoiseBound}
		\delta>\log (M)/M,
	\end{align}
	cf. Eq.~\eqref{gl:pM}.
	Note that for the limit $\delta\to\infty$, the distribution~\eqref{gl:pGauss} is a uniform distribution, in such a case $\Delta\varphi\to\infty$ and it is impossible to determine the phase at all.
	Consequently, the bound~\eqref{Eq:PhaseNoiseBound} limits the experimentally allowed amount of Gaussian phase noise.

\section{Application: Mixed N00N states and Propagation in turbulent media}\label{ch:App}
	Here we will study the entanglement and phase properties of mixed N00N states and N00N states after propagating in a turbulent quantum free-space link.
	In Sec.~\ref{ch:ModelAt} we described the model of the turbulent atmosphere focusing on the case of beam wandering in an urban environment.
	As an entanglement probe we will use the PT criterion.

\subsection{Partial transposition entanglement criterion}
	In this section we will briefly explain the PT entanglement criterion~\cite{Peres, Horodecki1996}.
	A bipartite state is entangled if it is not positive semidefinite after transposition of one mode.
	In this case it fails to be a density operator.
	The PT acts on the quantum state,
	\begin{align}
	   \hat\rho=\sum_{ijkl}p_{ijkl}|i\rangle \langle j|\otimes|k \rangle\langle l|
	\end{align}
	as
	\begin{align}
	   \hat\rho^{\rm PT}=\sum_{ijkl}p_{ijkl}|i\rangle \langle j|\otimes|l \rangle\langle k|.
	\end{align}
	If the state under study is negative under PT, $\hat{\rho}^{\rm PT}<0$, it has a minimal negative eigenvalue $\tau$ with a corresponding eigenvector $|\tau\rangle$.
	That is,
	\begin{align}\label{gl:PT}
	  \langle \tau|\hat{\rho}^{\mathrm{PT}}|\tau\rangle=\tau<0,
	\end{align}
	or, equivalently,
	\begin{align}
	  \mathrm{tr}(\hat{\rho}^{\mathrm{PT}}|\tau\rangle\langle\tau|)=\mathrm{tr}(\hat{\rho}[|\tau\rangle\langle\tau|]^{\mathrm{PT}})<0.
	\end{align}
	The negativity of $\tau$ is sufficient to verify entanglement.
	This method will be used to identify entanglement for the following scenarios. 
	
\subsection{Influence of mixing with vacuum noise}
	As a first example we aim to study both, entanglement and phase supersensitivity, under the influence of mixing.
	Therefore we study a mixture of a N00N and a vacuum state in the form
	\begin{align}\label{gl:N00Nvac}
	   \hat\rho_{p}=(1-p)|0,0\rangle\langle 0,0|+p|\psi_N\rangle\langle \psi_N| ,
	\end{align}
	with $p\in[0,1]$ and $|\psi_N\rangle$ being a N00N state~\eqref{gl:N00N} with $N>0$.
	The parameter $p$ controls the purity of this state with respect to an ideal N00N state.
	The mixture of a pure N00N state with vacuum is of particular interest, as it describes the transition from the N00N state to a phase-independent separable one.
	
\begin{figure}[h]
	\includegraphics[clip,width=1\linewidth]{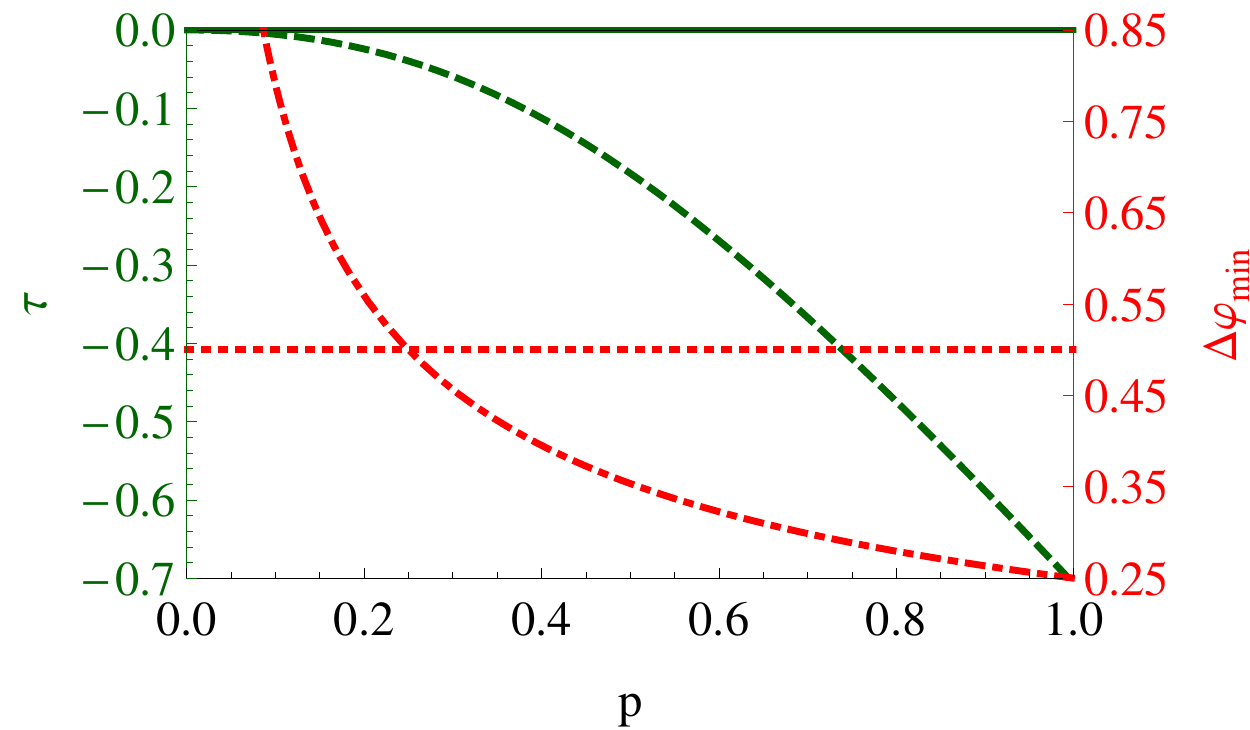}
	\caption{(Color online)
		Properties of the mixture~\eqref{gl:N00Nvac} of a N00N state with vacuum, for $N=4$, are shown in dependence on the noise parameter $p$.
		The minimal eigenvalue $\tau$ (dashed) [see Eq.~\eqref{Tau}] of the partial transposed state and the minimal phase error $\Delta\varphi_{\text{min}}$ (dash-dotted) [see Eq.~\eqref{gl:phasemixedN00N}] for $M=4$, are depicted on the left and right ordinates, respectively.
		The state is entangled if $\tau<0$, which is true for all nonzero $p$.
		Phase supersensitivity is achieved if the curve $\Delta\varphi_{\text{min}}$ is below the standard quantum limit of $1/\sqrt 4=0.5$ (dotted line).
	}
	\label{fig:EntP2}
	\end{figure}

	For this state one can apply the PT test and obtains that the state is entangled for all $p$ except $p=0$, corresponding to the vacuum state.
	Explicitly the minimal eigenvalue is 
	\begin{align}\label{Tau}
	   \tau=\frac{1-p-\sqrt{(1-p)^2+p^2}}{2},\\\nonumber
	   \text{with} \quad \tau<0 \quad\forall p\in]0,1].
	\end{align}
	In Fig.~\ref{fig:EntP2} this fact is illustrated in dependence on the mixing parameter $p$.
	This result shows that the entanglement property of the N00N state is very robust against mixing.

	Phase supersensitivity occurs for $p>0.25$; see Fig.~\ref{fig:EntP2}.
	However, in this case the bound, the standard quantum limit, and the minimal phase error, $\Delta\varphi_{\text{min}}=\min\{f_N(\varphi)/N\}$ [cf. Eq.~\eqref{gl:phasemixedN00N}] depend on the photon number $N$ of the N00N state.
	In general the $p$-interval for which phase supersensitivity exists for the state~\eqref{gl:N00Nvac} increases with increasing $N$.
	PT entanglement is shown to be present for every $p>0$ in Fig.~\ref{fig:EntP2}.

\subsection{Influence of turbulence}\label{sec:InfluTur}
	Now we will focus on the influence of turbulent losses on an initially pure N00N state.
	In particular we want to examine the entanglement properties of a N00N state including imperfections due to atmospheric beam wandering; see Sec.~\ref{ch:TurMixN00N}.
	Depending on the propagation distance, we study the properties of N00N states in urban quantum free-space links for the parameters $W_0=0.98$ mm, $C_n=10^{-17}$ m${}^{-2/3}$, and $a=W$ (see also Appendix \ref{app:Weinbull}).
	The choice of parameters is done according to the experimental realization in~\cite{Heim,Elser}.
	This set of parameters corresponds to a mean transmission coefficient of $\overline T\approx0.843$ at a distance of $200$ m within one mode.
	For the entanglement verification, we again will apply the partial transposition criterion and consider its corresponding smallest eigenvalue $\tau$ in dependence on the propagation distance.

\begin{figure}[h]
	\includegraphics[clip,width=1\linewidth]{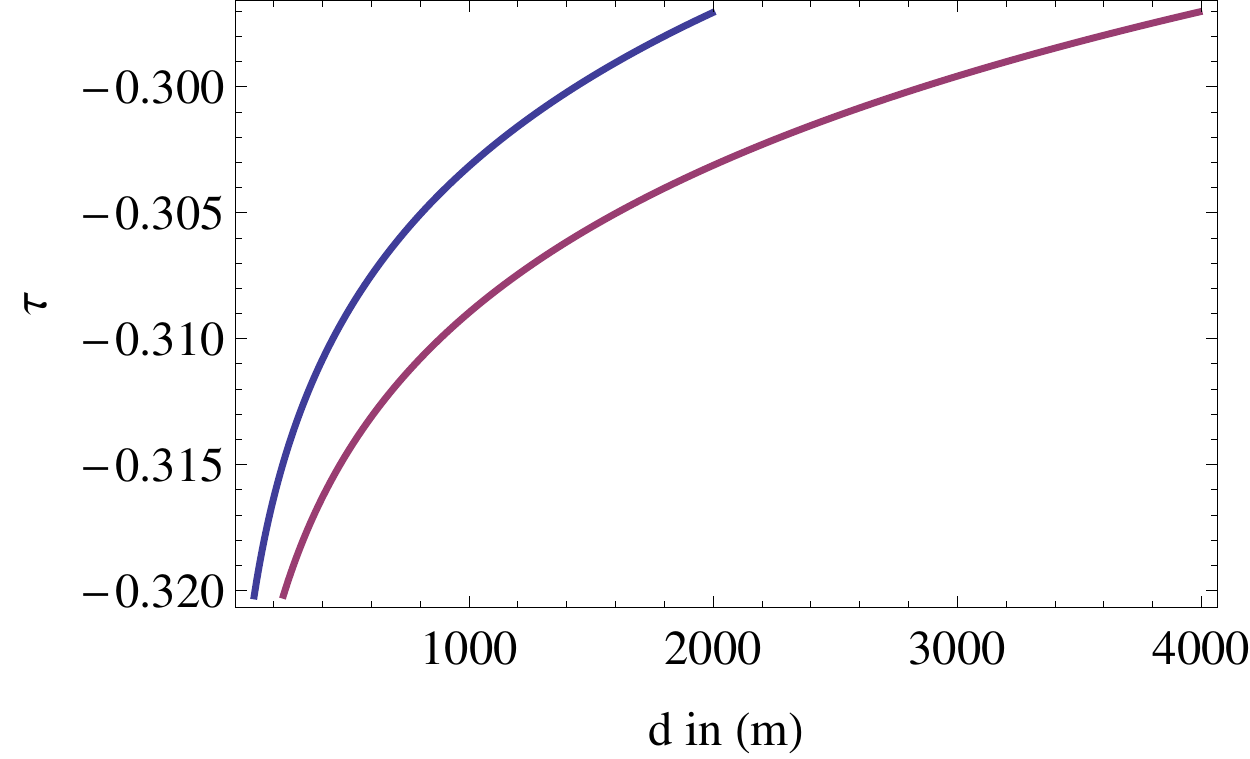}
	\caption{(Color online) 
		The entanglement test is shown for a turbulent N00N state with $N=2$ for copropagation (left curve) and counterpropagation (right curve).
		The minimal eigenvalue $\tau$ is displayed in dependence on $d$, representing the distance between sender and receiver (left curve) and the distance between the two receivers (right curve).
		Entanglement can be preserved in such a turbulent environment, without any postselection over distances of about $2000$ m and $4000$ m for the co and counterpropagation, respectively.
	}
	\label{fig:TurLoss}
	\end{figure}

	In Fig.~\ref{fig:TurLoss} this entanglement test is displayed for $N=2$ in the counterpropagation and copropagating cases [see Eqs.~\eqref{eq:counter-prop} and~\eqref{eq:co-prop}, respectively].
	Its dependency on $d$, the distance between the two receivers and the sender and receiver for counter and copropagation, respectively,  is shown.
	Here one sees that entanglement of N00N states can survive in such an urban link and entanglement can be transferred up to a distance of about $4000$ m.
	In principle the entanglement test would identify entanglement for any propagation distance; however the PDTC model used would not hold true for significantly larger distances.
	Note that other PDTC models, which would describe such other atmospheric scenarios, are not known yet and are subject of current research.
	Additionally it is important to stress that typically applied postselection protocols may improve the range of successful entanglement propagation significantly; see, e.g., the approaches in Refs.~\cite{Ursin,Fedrizzi,Semenov2010,beamwandering}.

	A substandard quantum limit behavior of phase estimation cannot be observed for the distances $d$ in Fig.~\ref{fig:TurLoss}, since phase super-sensitivity of noisy N00N states is obtained only if the matrix entries $\rho_{0N,N0}$ are large enough [see Eq.~\eqref{gl:pM}].
	However, they scale with the $2N$th moment of the corresponding PDTC distribution.
	More generally, each interference term after an atmospheric channel is rescaled with
	\begin{align}
		\int\limits_0^{1}\int\limits_0^{1}d\sqrt{\kappa}\,d\sqrt{\theta}\,
		\mathcal{P}(\sqrt{\kappa},\sqrt{\theta}) \sqrt{\kappa\theta}^N,
	\end{align}
	see Eqs.~\eqref{gl:lossy} and~\eqref{eq:PDTCchannel}.
	Since $\kappa,\theta\leq 1$, we get a decreasing interference term for increasing $N$ and for every PDTC--$\mathcal P(\sqrt{\kappa},\sqrt{\theta})$ -- except in the ideal case of no loss, i.e., $\mathcal{P}(\sqrt{\kappa},\sqrt{\theta})=\delta(\sqrt{\kappa}-1)\delta(\sqrt{\theta}-1)$.

	Let us stress again that the coherences $\rho_{0N,N0}$ are crucial for determining both the phase supersensitivity and the entanglement.
	On the one hand, a small value of $N$ might be preferable in some quantum communication scenarios in turbulent media.
	On the other hand, a large $N$ value is favorable in the non-perturbed case.
	Hence, the present method can be used to predict an optimal choice of $N$ conditioned on the actual turbulence properties of a free-space link.
	This example also demonstrates that entanglement of N00N states tolerates turbulent losses better than its phase supersensitivity property.

\section{Summary and conclusions}\label{ch:Summary}
	Motivated by the stochastic generation of N00N states and by imperfections such as losses and dephasing, we introduced a generalized class of noisy N00N states.
	For this class of states conditions for sub-standard quantum limit behavior in a Mach-Zehnder interferometer have been studied depending on certain perturbations of N00N states.
	In particular, the influence of dephasing on phase supersensitivity has been considered.

	The propagation of noisy N00N states in atmospheric links--described as fluctuating loss channels--has been investigated.
	Here we could identify, for the effect of beam wandering, certain regions of turbulence which partly preserve the phase supersensitivity of the perfect N00N states.
	Applying the partial transposition condition, we could certify that entanglement is preserved in an urban quantum free-space link.

	In conclusion, our approach allows the verification of quantum properties of light in highly perturbed systems.
	The described methods yield lower bounds for the survival of quantum features after propagation in free-space communication links.
	The bounds might be enhanced by employing postselection strategies.
	Our technique yields a useful tool for experimentalists to predict the preservation or loss of quantum correlations in terms of entanglement and phase supersensitivity.
	Hence, the outlined method opens the possibility of employing perturbed quantum states in authentic applications, such as high precision measurements in quantum metrology and quantum key distribution for secure quantum communication.

\begin{acknowledgments}
	This work was supported by the Deutsche Forschungsgemeinschaft through Grant No. VO 501/22-1 and SFB 652 (B2).
\end{acknowledgments}

\appendix

\section{Parameters of the atmosphere}\label{app:Weinbull}
	In Eq.~\eqref{gl:tur} the log-negative Weibull distribution was given which describes the PDTC of beam wandering.
	The needed parameters are~\cite{beamwandering}
	\begin{align}
		W=&\,W_0\sqrt{1+\left(\frac{z\lambda}{\pi W_0^2}\right)^2}\\
		T_0^2=&\,1-\exp\left[-2\frac{a^2}{W^2}\right]\;,\\ \nonumber
		\zeta =&\,8\frac{a^2}{W^2}\frac{\exp\left[-4\frac{a^2}{W^2}\right]I_1\left(4\frac{a^2}{W^2}\right)}{1-\exp\left[-4\frac{a^2}{W^2}\right]I_0\left(4\frac{a^2}{W^2}\right)} \\ 
		&\times \left[\ln\left(\frac{2T_0^2}{1-\exp\left[-4\frac{a^2}{W^2}\right]I_0\left(4\frac{a^2}{W^2}\right)}\right)\right]^{-1}\;,\\
		S=&\,a\left[\ln\left(\frac{2T_0^2}{1-\exp\left[-4\frac{a^2}{W^2}\right]I_0\left(4\frac{a^2}{W^2}\right)}\right)\right]^{-\frac{1}{\zeta}}\\
		\intertext{and}
		\sigma^2 \approx & 1.919\,C_n^2z^3(2W_0)^{-\frac{1}{3}}\;.
	\end{align}
	The meaning of the remaining parameters are listed in Table~\ref{tab:Coef}.
	\begin{table}[ht!]
	  \caption{Coefficients of the PDTC.}
	  \label{tab:Coef}
	  \centering
	  \begin{center}
	    \begin{tabular}{ c  l }
	      \hline\hline
	      Symbol & Meaning\\ \hline
	      $W$ & beam-spot radius at the aperture \\ 
	      $W_0$ & beam-spot radius at the radiation source \\ 
	      $a$ & aperture radius of the detector \\ 
	      $z$ & distance between radiation source and detector\\ 
	      $S$ & scale parameter\\ 
	      $\zeta$ & shape parameter \\ 
	      $\sigma^2$ & variance of the beam-center position \\ 
	      $C_n^2$ & index of refraction structure constant \\ 
	      $\lambda$ & wavelength \\ 
	      $I_n$ & $n$th modified Bessel function \\ \hline\hline
	    \end{tabular}
	  \end{center}
	\end{table}


\end{document}